\documentclass{elsart3p}
\usepackage{graphicx,amssymb}
\usepackage[square,numbers]{natbib}
\begin{document}
\begin{frontmatter}

\title{Study of the internal mechanisms of
Pixelized Photon Detectors operated in Geiger-mode}
\author[Tokyo]{H.~Otono\corauthref{cor}},~ 
\corauth[cor]{corresponding author.}
\ead{otono@icepp.s.u-tokyo.ac.jp}
\author[Tokyo]{H.~Oide},~ 
\author[ICEPP]{S.~Yamashita},~
\author[KEK]{T.~Yoshioka},
\author[HPK]{K.~Yamamoto},~
\author[HPK]{K.~Yamamura},~
\author[HPK]{K.~Sato}

\address[Tokyo]{Department of Physics, School of Science, The University of Tokyo, 7-3-1 Hongo, Bunkyo-ku, Tokyo 113-0033, Japan}
\address[ICEPP]{International Center for Elementary Particle Physics, The University of Tokyo, 7-3-1 Hongo, Bunkyo-ku, Tokyo 113-0033, Japan}
\address[KEK]{Neutron Science Laboratory, High Energy Accelerator
Research Organization (KEK),
1-1 Oho, Tsukuba, Ibaraki 305-0801, Japan}
\address[HPK]{Solid State Division, Hamamatsu Photonics K.K.,
1126-1 Ichino, Higashi-ku, Hamamatsu, Shizuoka 435-8558, Japan}
\begin{abstract}
  In the 1990s, a novel semiconductor photon-sensor operated in Geiger-mode
  was invented in Russia (Silicon PhotoMultiplier),
  which consists of many tiny pixels and has a single photon level sensitivity. 
  Since then, various types of the sensor with this scheme, 
  Pixelized Photon Detectors (PPD), have been developed 
  in many places in the world.
  For instance, Hamamatsu Photonics K.K. in Japan produces 
  the PPD as a Multi-Pixel Photon Counter.  
    
  While the internal mechanisms of the PPD have been intensively studied 
  in recent years, the existing models do not succeeded to fully 
  reproduce the output characteristic, such as waveforms at low temperature.
  We have developed a new model with the transient multiplication and quenching
  of the realistic avalanche process 
  and have succeeded in reproducing the output 
  waveform of the PPD at various temperature.
  In this paper, we discuss our improved model.
\end{abstract}

\end{frontmatter}

\section{Introduction}
In the 1990s, a novel semiconductor photon-sensor operated in Geiger-mode
was invented in Russia (Silicon PhotoMultiplier) 
\cite{Golovin:1998,Sadygov:1998},
which consists of many tiny pixels and has a single photon level sensitivity. 
Since then, various types of the sensor with this scheme
have been developed in many places in the world 
\cite{Renker:2006ay}. 
For instance, Hamamatsu Photonics K.K.
(HPK) in Japan produces a Multi-Pixel Photon Counter (MPPC) \cite{Yamamoto:2007PD07,Gomi:2007zz}.  
Recently these devices are called 
as Pixelized Photon Detectors (PPD) \cite{Amsler:2008zz}.

Since the PPD has many advantages over PMTs, such as compact size, low cost,
low power consumption and insensitive to magnetic field, 
it could be a successor to them in high energy 
experiments and/or in medical fields in the near future.

The particular device studied in this paper is S10262-11-025 of MPPC by HPK
which has a ${\rm 1mm\times1mm}$ photosensitive surface and 
consists of 1600 pixels in total.

We studied the basic properties of the MPPC at low temperature
\cite{Otono:2007PD07} and found that
output waveform has a spike component which can not be fully explained
by existing models. 
We introduced a new model in which the realistic avalanche process is considered
and succeeded in reproducing the output waveform at various temperature.

  \section{Existing models of the PPD}
  \subsection{\rm{The traditional model}}

\begin{figure}[h]
  \begin{center}
   \includegraphics*[width=75mm]{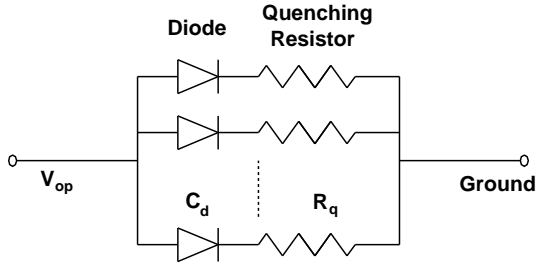}
  \end{center}
  \caption{A simple circuit of the PPD. Each pixel consists
  of a diode and a quenching resistor.}
  \label{fig:circuit}
 \end{figure}

  Figure \ref{fig:circuit} shows the traditionally used equivalent circuit for
  the PPD which explains 
  the linearity between the gain ($G$) and the operating voltage ($V_{op}$).
  The PPD has multiple pixels and each pixel consists of a diode and 
  a quenching resistor ($R_q$).
  The PPD is operated at a few volts higher than 
  the breakdown voltage ($V_{br}$), namely Geiger-mode, 
  and the diode itself does not have the 
  capability to terminate the avalanche process. 
  Hence, the quenching resistor 
  is necessary in order to decrease the voltage applied to the diode. 
  When the current fed into the resistor is
  $I_{\rm{expected}}\equiv \frac{\Delta V}{R_{q}}$,
  where $\Delta V$ is $V_{op}-V_{br}$, 
  the voltage applied to
  the diode drops to $V_{br}$ and the multiplication stops.
  As a result, the gain is obtained as follows: 
  $G=\frac{C_d \times \Delta V}{e}$,
  where $C_d$ is the capacitance value of the diode and 
  $e$ the elementary electric charge. 
  
\subsection{\rm{Waveform at low temperature}}

\begin{figure}[h]
  \begin{center}
    \includegraphics*[width=85mm]{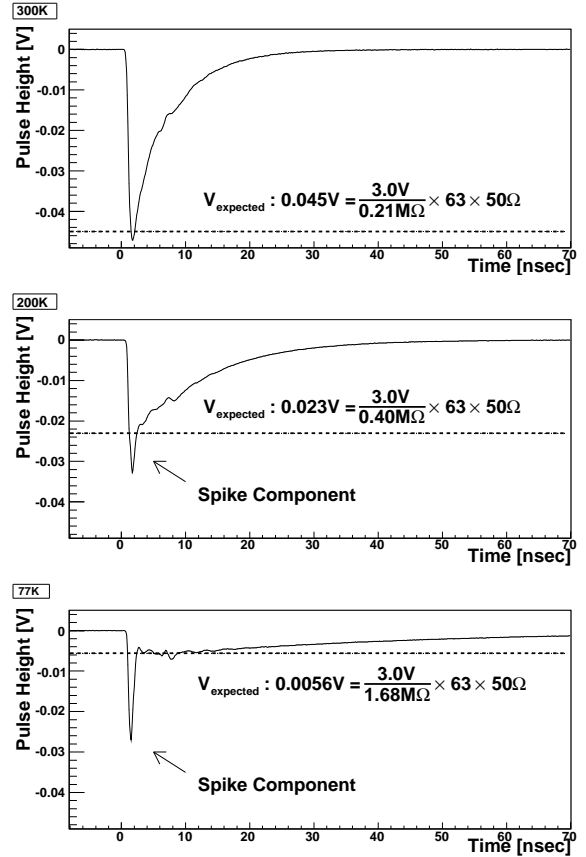}
  \end{center}
  \caption{Waveforms observed at 300K (top), 200K (middle) and
    77K (bottom). The waveforms at 200K and 77K have two components.}
  \label{fig:wave}
\end{figure}

We studied the basic properties, such as waveforms, of 
MPPC produced by HPK at 300K, 200K and 77K \cite{Otono:2007PD07}.
The used amplifier was C5549 (bandwidth 1.5GHz) produced by HPK
whose multiplication factor and input impedance are 63 and 50$\Omega$, 
respectively.
Figure \ref{fig:wave} shows the average waveform at each temperature,
where $\Delta V$ is fixed at 3.0V, and
a spike component is seen in the waveforms taken at 200K and 77K.
Although the existence of two components has already been reported 
by the ITC-irst group at room temperature with a SiPM 
\cite{Piemonte:2007IEEE,Piemonte:2007NIMA}, 
it is the first time to observe the sharp spike component shown 
in the waveform of MPPC at 77K.

$R_q$ can be measured from the forward
current-voltage characteristics of the PPD by using the following equation:
$\frac{R_q}{N_{\rm{pixels}}}=\frac{V_{f}}{I_{f}}$, 
where $I_{f}$,$V_{f}$ and $N_{\rm{pixels}}$ are the forward current value, 
the forward voltage and the number of pixels, respectively.
The resistor values obtained were shown in Table \ref{tb:list}.

The negative temperature coefficient of resistivity is due to
the amorphous-like structure of poly-silicon \cite{Kato:1982IEEE}
which is the main component of the quenching resistor in the MPPC.

\subsection{\rm{A problem on the traditional model}}
Dashed lines in the Figure \ref{fig:wave} indicate the expected pulse height ($V_{\rm{expected}}$) by the traditional model, 
which can be obtained from the following equation:
$V_{\rm{expected}}=I_{\rm{expected}}\times A \times Z_{\rm{input}}$,
where $A$ is the multiplication factor of the amplifier and $Z_{\rm{input}}$ 
is the input impedance.

The spike component becomes much higher than $V_{\rm{expected}}$ 
at low temperature.
The traditional model is thus not sufficient for explaining the output
waveform of the PPD.      
$V_{\rm{expected}}$ and the measured pulse height ($V_{\rm{measured}}$) at
each temperature are also summarized in Table \ref{tb:list}.

\begin{table}[htbp]
  \begin{center}
    \begin{tabular}{cccc}
      \hline      
      $T$[K] & $R_{q}$[M$\Omega$] & $V_{\rm{expected}}$[V] & $V_{\rm{measured}}$[V] \\ 
      \hline
      300    &  0.21              &  0.045               &  0.047  \\
      200    &  0.40              &  0.023               &  0.033  \\
      77     &  1.68              &  0.0056              &  0.027  \\  
      \hline  \\ 
    \end{tabular}   
    \caption{$R_q$, $V_{\rm{expected}}$ and $V_{\rm{measured}}$ at
each temperature.} 
    \label{tb:list}        
  \end{center}
\end{table}

\subsection{\rm{A model developed by the ITC-irst group}}
\begin{figure}[h]
  \begin{center}
    \includegraphics*[width=75mm]{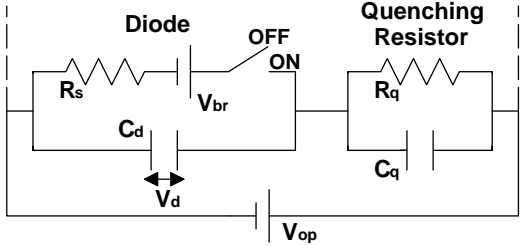}
  \end{center}
  \caption{An equivalent circuit of a single pixel which was developed by the 
ITC-irst group \cite{Piemonte:2007IEEE,Piemonte:2007NIMA}.}
  \label{fig:circuit2}
\end{figure}

ITC-irst group looked into a capacitance ($C_q$) located 
between the quenching resistor 
and the diode as a possible reason for explaining two components. 
$C_q$ is typically in the order of 1fF \cite{Corsi:2007NIMA}.
Figure \ref{fig:circuit2} shows its equivalent circuit proposed by them.
They simplified the scheme of the multiplication and quenching
and substituted a switch. 
$R_{s}$ is the series resistance of the diode.

As a consequence, they succeeded in reproducing the two components 
by tuning the parameters, such as $R_s$ and $C_q$.

\section{A new model including the avalanche process}
Although the existence of the two components is explained by
ITC-irst group, the avalanche process is described by the switch and
the transient multiplication is not included in their model.
We introduce a new model in which the realistic 
avalanche process is considered so as to reproduce well the output waveform.

\subsection{\rm{A reproduction of the output waveforms}}
\begin{figure}[h]
  \begin{center}
    \includegraphics*[width=75mm]{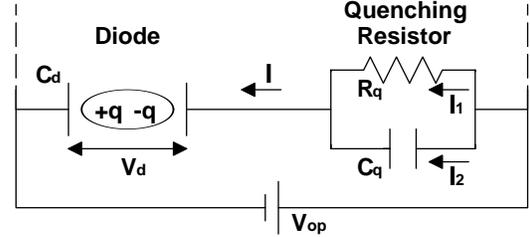}
  \end{center}
  \caption{A new equivalent circuit proposed in this paper.}
  \label{fig:circuit3}
\end{figure}

Our model is shown in Figure \ref{fig:circuit3} and 
it is described by the charge of created carriers $q(t)$ in the multiplication layer in the diode
rather than by the switches of the model by ITC-irst group.

$\frac{dq(t)}{dt}$ depends on the ionization rate of electrons and holes, $\alpha_e$ and $\alpha_h$, 
which are proportional to ${E(x)}\exp(-\frac{B}{E(x)})$, where $E(x)$ is the electric field strength and $B$ is a constant \cite{Maes:1990Solid}.
We calculated the ionization rate from the interpolated values of the measurement by Grant \cite{Grant:1973Solid}.
Thus $\frac{dq(t)}{dt}$ can be expressed by: 

\begin{displaymath}
  \left \{
  \begin{array}{l}
    \frac{d q(t)}{dt}=\sum_{i=e,h}\int e{\rho_i}_{(x)}{\alpha_i}_{(E_{(x)})} {v_i}_{(E_{(x)})}
~dx^3\\
    V_d(t)=\int^L_{0}E(x)~dx,
  \end{array}
  \right.
\end{displaymath}
where $\rho_e$ and $\rho_h$ are the densities
of electrons and holes in the diode,
$v_e$ and $v_h$ are the drift velocities of each carrier,
$V_d(t)$ is the voltage applied to the diode and
$L$ is the width of the diode.
In this calculation, $L$ is fixed at $2\mu m$. 

We fixed $v_e$ and $v_h$ at $v=1\times10^7\rm{cm/sec}$, because
they are saturated and remain constant 
in the high electric field at any temperature \cite{Jacobi:2007Solid}, 
and ignored the temperature dependence of $\alpha_e$ and $\alpha_h$ \cite{Rose:1998APPL}.
We calculated the electric field of the typical PPD \cite{Kagawa:2005IEICE}
and obtained the almost uniform electric field in the multiplication layer in the diode,
thus we substitute $\alpha_{({V_d}_{(t)})}$ for $\alpha_{({E}_{(t)})}$.
As a result, the equations are simplified as follows:
\begin{equation}
  \frac{d q(t)}{dt}=ev\sum_{i=e,h}\int{\rho_i}_{(x)}{\alpha_i}_{({V_d}_{(t)})}~dx^3.
  \label{eqn:dq}
\end{equation}

Next, the circuit equations are the following:

\begin{displaymath}
  \left \{
  \begin{array}{l}
    I=I_1+I_2\\
    V_{op}=V_d(t)+I_1 R_q\\
    V_{op}=V_d(t)+\frac{\int_{0}^{t}I_2~dt}{C_q},
   \end{array}
  \right.
\end{displaymath}
where $I$ is the observable current.
The equation of equilibrium of the charge in the diode is 
calculated from $q(t)$ and $I(t)$:
\begin{displaymath}
  {C_d}\frac{dV_d(t)}{dt}=\frac{dq(t)}{dt}-I(t).
\end{displaymath}
$I_1$, $I_2$ and $I$ are substituted into the first equation
and the following differential equation can be obtained:

\begin{equation}
  \frac{d V_d(t)}{dt}=\frac{1}{R_q(C_d+C_q)}(V_{op}-V_d(t)-R_q\frac{dq(t)}{dt}).\\
  \label{eqn:dv}
\end{equation}

From Eqs.(\ref{eqn:dq}) and (\ref{eqn:dv}), $V_d(t)$ and $q(t)$ 
are sequentially solved and the output current of the PPD can be
expressed by:

\begin{displaymath}
I(t)=C_d \frac{d V_d(t)}{dt}+\frac{d q(t)}{dt}.
\end{displaymath}

\begin{figure}[h]
  \begin{center}
    \includegraphics*[width=85mm]{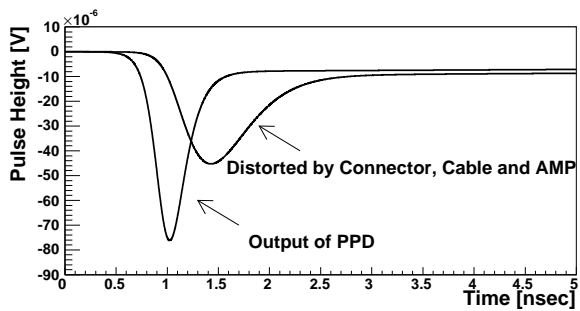}
  \end{center}
  \caption{Expected output of the PPD and the distortion due to connectors, cables and an amplifier.}
  \label{fig:PulseDistortion}
\end{figure}

For the purpose of comparison with observed current, 
the frequency characteristics of the used connectors, cables and amplifier 
have to be taken into account \cite{Brianti:1965CERN}.
For our system the transmission properties are determined by the connector
since it has the smallest bandwidth of all components at 1.4GHz (Suhner QLA).
The bandwidths of the amplifier and the cable are 1.5GHz and
2.0GHz, respectively. 
Figure \ref{fig:PulseDistortion} shows the pulse distortion of our system.

Hence, the leading and trailing edge of the spike component at 77K are 
characterized by the bandwidth of the connector, 1.4GHz.
$R_q$ is already obtained as 1.68M$\rm{\Omega}$ and
$C_d$ can be calculated from the time constant of the recovery process, 
$R_qC_d$ \cite{Oide:2007PD07}.
In order to calibrate $C_q$, we evaluate the pulse height 
of spike component at 77K, because it is the most sensitive to the variation of $C_q$. 
Consequently, we can obtain $C_q$ as 2fF.

\begin{figure}[h]
  \begin{center}
    \includegraphics*[width=85mm]{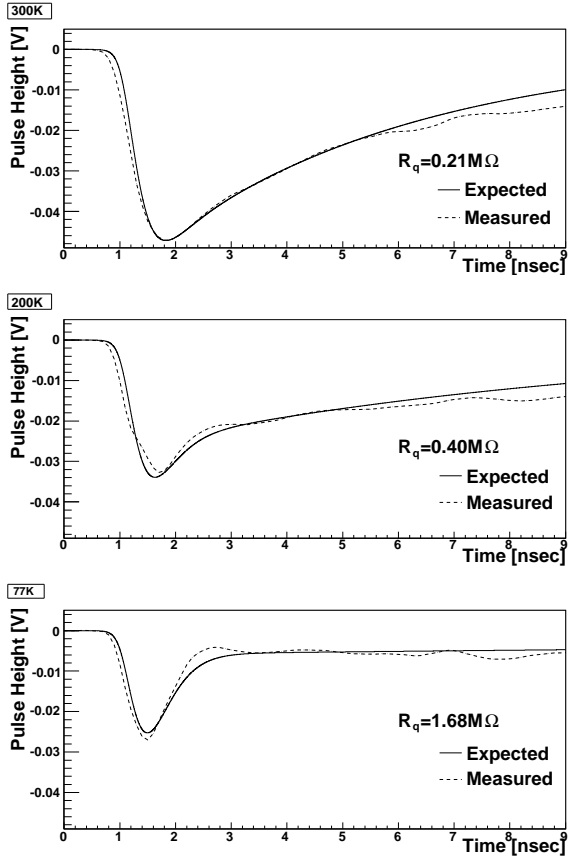}
  \end{center}
  \caption{Closeup view of Figure \ref{fig:wave}. The waveform expected from our proposed model and the measured 
waveform.}
  \label{fig:Final}
\end{figure}

The expected and measured waveforms at each temperature
are shown in Figure \ref{fig:Final} by the solid and dashed curves, 
respectively.
All parameters are summarized in Table \ref{tb:list2}.
Note that $C_d$ and $C_q$
are calibrated from the waveform at 77K.
The same values are applied to the waveform at 200K and 300K. 
\begin{table}[htbp]
  \begin{center}
    \begin{tabular}{llll}
      \hline      
      $T$[K] & $R_{q}$[M$\Omega$] & $C_d$[fF] & $C_q$[fF] \\ 
      \hline
      77     &  1.68       &  20.2     &  2 \\  
      300    &  0.21       &  20.2     &  2  \\
      200    &  0.40       &  20.2     &  2 \\
      \hline  \\ 
    \end{tabular}   
    \caption{$R_q$ at each temperature is measured value. $C_d$ and $C_q$
      are calibrated from waveform at 77K.
      The same values are applied to the waveform at 200K and 300K.} 
    \label{tb:list2}        
  \end{center}
\end{table}

The waveforms in  Figure \ref{fig:Final} are the closeup view of 
Figure \ref{fig:wave} and
the expected waveforms compare well with the measured ones. 
Therefore, our model properly includes the multiplication and quenching process 
and reproduces the output waveform of the PPD.

\subsection{\rm{A reproduction of $G-\Delta V$ relationship}}
\begin{figure}[h]
   \begin{center}
     \includegraphics*[width=85mm]{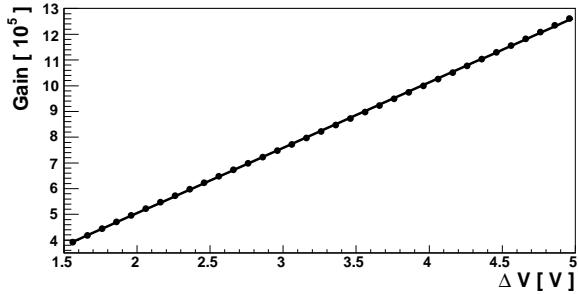}
   \end{center}
   \caption{The linearity between the gain and the operation
     voltage, which is calculated from our model.}
   \label{fig:Linear}
 \end{figure}

In the model by ITC-irst group, 
the linearity between the gain of the PPD ($G$)
and $\Delta V$ is determined by the opening of the switch when $V_d$ drops to
$V_{br}$, which forces the termination of the avalanche process.
This guarantees that the gain relationship is in accord with PPD 
measurements, for which the relationship is indeed linear.
In our model there is no such forcing, and as such $V_d$ drops below
$V_{br}$.
Thus, this linearity is not trivially determined.
We have though calculated the $G-\Delta V$ relationship from the model and
have in fact found it to be linear as shown in Figure \ref{fig:Linear} 
which is further verification of the model.

\section{Conclusion}

We developed a new model for a Pixelized Photon Detector (PPD),
which includes the dynamical avalanche process.
By calculating the circuit equations of the model, 
the multiplication 
and quenching process can be naturally incorporated as following.

The moment the avalanche process occurs, the current feeds into 
the capacitance $C_q$ and is observed as the spike component of the waveform.
The leading and trailing edge of the spike component is determined by the transmission properties of the measurement system. 
At the same time, the avalanche multiplication begins to be suppressed 
by the current fed into the quenching resistor $R_q$
and finally quenches.
After quenching, the recovery process is characterized by the time constant,
$R_qC_d$, where $C_d$ is the capacitance value of the diode.

Furthermore we have succeeded in reproducing the observed characteristic 
waveforms, such as sharp spike component at 77K.
The waveforms at 200K and 300K can also reproduced with 
$C_d$ and $C_q$ calibrated from the waveform at 77K and measured $R_q$.
Note that there is no variable for reproducing the waveforms at 200K and 300K.

Consequently, the model proposed in this paper makes it possible to properly
incorporate the internal mechanisms of the PPD.
Therefore, it should play an important role for the
development of PPDs in the future.

\section{Acknowledgments}
The authors wish to express our deep appreciation to the KEK-DTP
photon-sensor group members for their helpful discussions and suggestions.
This work was supported by Grant-in-Aid for JSPS Fellows $20\cdot4439$
and by Grant-in-Aid for Exploratory Research $20654021$
from the Japan Society for the Promotion of Science (JSPS).


\end{document}